\documentclass[twocolumn]{aastex631}

\submitjournal{ApJ}
\usepackage{cuted}
\usepackage{amsmath}
\usepackage{booktabs}
\usepackage{float}
\usepackage{longtable}
\usepackage{cuted}

\begin{document}
\title{Polarimetric Signatures of Bulk Comptonization from within the Plunging Region of Accreting Black Holes}

\author[0000-0003-2776-082X]{Ho-Sang Chan}
\email{hschanastrophy1997@gmail.com}
\altaffiliation{Croucher Scholar}
\affiliation{JILA, University of Colorado and National Institute of Standards and Technology, 440 UCB, Boulder, CO 80309-0440, USA}
\affiliation{Department of Astrophysical and Planetary Sciences, University of Colorado, 391 UCB, Boulder, CO 80309, USA}

\author[0000-0003-0936-8488]{Mitchell C. Begelman}
\affiliation{JILA, University of Colorado and National Institute of Standards and Technology, 440 UCB, Boulder, CO 80309-0440, USA}
\affiliation{Department of Astrophysical and Planetary Sciences, University of Colorado, 391 UCB, Boulder, CO 80309, USA}

\author[0000-0003-3903-0373]{Jason Dexter}
\affiliation{JILA, University of Colorado and National Institute of Standards and Technology, 440 UCB, Boulder, CO 80309-0440, USA}
\affiliation{Department of Astrophysical and Planetary Sciences, University of Colorado, 391 UCB, Boulder, CO 80309, USA}


\begin{abstract}

Inverse Compton scattering by the thermal motions of electrons is believed to produce polarized hard X-rays in active galactic nuclei and black-hole binaries. Meanwhile, plasma within the plunging region of the black hole free falls into the event horizon with a bulk relativistic speed, which could also imprint polarization on up-scattered photons but has not been discussed in detail. To examine this, we computed polarimetric signatures via general relativistic ray-tracing of a toy model consisting of an accreting, geometrically thin plasma with moderate optical depth, falling onto the black hole with a bulk relativistic speed within the plunging region. We show that the maximum spatially unresolved linear polarization could be as large as approximately $7 - 8$\,\% when the black hole is viewed near edge-on, while the corresponding resolved linear polarization could be roughly $50$\,\%. The large discrepancy between the two is due to 1) dilution from the radiation outside the plunging region and 2) substantial cancellations of the Stokes $Q$ and $U$ fluxes. The resultant polarization contributed by bulk Comptonization could nevertheless exceed that of thermal electron scattering in a Novikov–Thorne disk. Our results thus suggest a new model for imprinting considerable polarization on the electromagnetic observables of accreting black holes. Measurements of X-ray polarization from black-hole binaries and the central black hole of active galactic nuclei could provide direct detection of the plunging region and help constrain plasma properties in the immediate vicinity of the event horizon.

\end{abstract}



\section{Introduction} \label{sec:intro}


The study of electromagnetic signatures from accreting black holes could reveal details about plasma properties \citep{chan2015power, chan2023230, chan2024230, grigorian2024relationship}, spacetime metrics \citep{akiyama2022first}, and magnetic field structures around black holes \citep{Akiyama_2021}. For instance, inverse Compton scattering — whereby photons gain higher energies by colliding with high-speed electrons — is thought to be crucial in creating the observed spectra of accreting black holes and in imprinting polarization. In most models, the motions of the scattering electrons are random, with velocity distributions that may be purely thermal or include a nonthermal tail \citep{mcconnell2002soft, schnittman2010x, hankla2022non}.

Here, we are interested in the bulk Comptonization of photons from within the plunging region (or innermost stable circular orbit (ISCO), which we will use interchangeably). The motivation is that there is no stable circular orbit within the ISCO, and plasma will inevitably free-fall towards the event horizon with a bulk relativistic speed. Meanwhile, semi-analytic, thin accretion disk solutions \citep{hankla2022non, mummery2024continuum} and numerical simulations of black-hole binaries (BHB) in the hard state \citep{dexter2021radiation} indicate a substantial plasma density drop within the ISCO, making the plasma marginally optically thick ($\sim O(1)$). Plasma within the ISCO thus 1) has considerable bulk speed and 2) is marginally optically thick. Therefore, the plunging region should be an ideal site for bulk Comptonization to operate. Predicting the electromagnetic observables from within the ISCO has gained interest in recent years \citep{wilkins2020venturing, potter2021full, mummery2024continuum}. However, the polarized emission from the plunging region has hardly been explored; predicting it could help constrain the plasma properties within that region. This is especially true in light of the recent launch of the \textit{Imaging X-ray Polarimetry Explorer} (\textit{IXPE}), which is  providing valuable information about X-ray spectra and linear polarization in the $2 - 8$\,keV range, e.g., in the BHBs Cygnus X-1 \citep{krawczynski2022polarized} and 4U 1630–47 \citep{kushwaha2023ixpe}. 

In this paper, we extend the formalism in \citet{begelman1987inverse} and \citet{dexter2024relativistic} to examine how bulk Comptonization by relativistically moving plasma within the plunging region affects the polarimetry of accreting black holes. This paper is structured as follows: In Section \ref{sec:method}, we present an overview of our toy accreting plasma model, review the physics of bulk Comptonization, and demonstrate how to compute the observed polarization using a general relativistic ray-tracing (GRRT) code. In Section \ref{sec:results}, we present our results, highlighting how the polarimetry depends on the model parameters. As we shall see, there is a significant discrepancy between the spatially resolved and unresolved linear polarization, which arises from 1) dilution by radiation from the disk and 2) the substantial cancellation of alternating-sign Stokes $Q$ and $U$ fluxes. Nonetheless, the maximum linear polarization could reach up to approximately $7 - 8$\,\% when the black hole is viewed near edge-on, even with these effects taken into account. In Section \ref{sec:discuss}, we discuss the implications of our results, and we conclude our study in Section \ref{sec:conclu}.


\section{Methodology} \label{sec:method}


%
\begin{figure*}[htb!]
    \centering
    \includegraphics[width=1.0\linewidth]{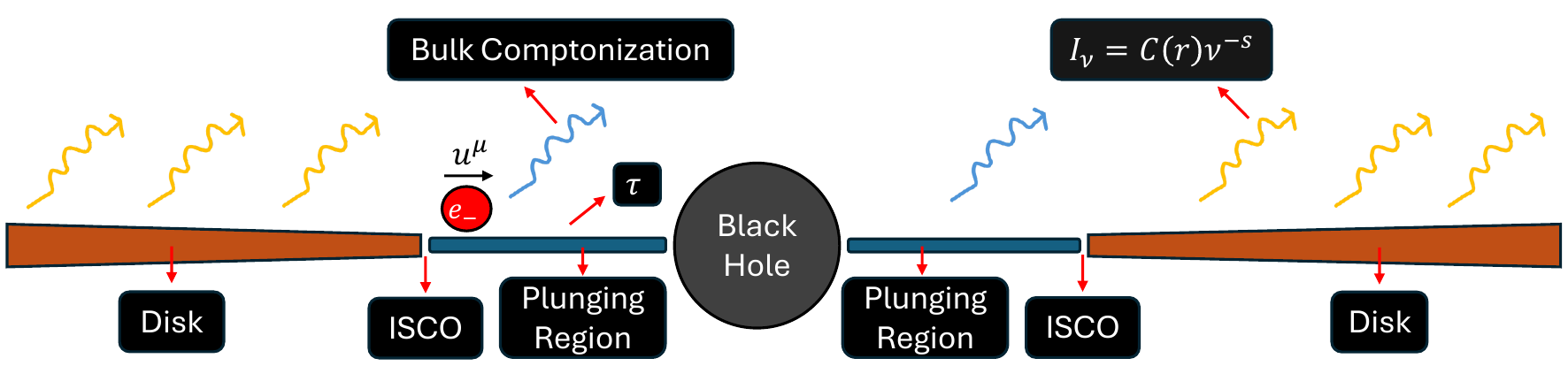}
    \caption{The setup of our toy plasma model. The plasma inside the ISCO is geometrically thin, has a bulk velocity (described by the bulk 4-vector $u^{\mu}$, assuming spherical Boyer-Lindquist coordinates), and is characterized by the black hole spin $a$, the plunging region optical depth $\tau$, and the background intensity $I_{\nu}$. We assume that $I_{\nu}$ follows a power-law spectrum, and the proportionality constant $C(r)$ is a function of the radius $r$ from the black hole. The intensity $I_{\nu}$ is reprocessed by bulk Comptonization within the ISCO only. \label{fig:setup}}
\end{figure*}

Here, we provide an overview of the toy plasma model used in our study, as illustrated in Figure \ref{fig:setup}. We assume the presence of accreting plasma around the black hole, which is geometrically thin. The background radiation field\footnote{contributed by local plasma emission and/or an external radiation source} follows a power-law spectrum, with the proportionality constant being a function of the radius $r$ from the center of the black hole. Inside the ISCO, the background radiation field is reprocessed by bulk Comptonization and imprinted with polarization. Beyond the ISCO, the optical depth of the plasma could become so large that the single-scattering approximation of the bulk Comptonization model would fail. Therefore, we do not include any bulk Comptonization physics outside the ISCO. The accreting plasma inside the ISCO has a bulk $4-$velocity, $u^{\mu}$, and is characterized by three parameters: the black hole spin, the radiation profile, and the (assumed constant) plunging region optical depth. This study aims to examine how the variation of these parameters affects the polarization signatures of the accreting plasma.


\subsection{Recap of Bulk Comptonization} \label{subsec:review}


To study how bulk Comptonization reprocesses the background radiation field and imprints polarization within the ISCO, we first briefly review the formalism of bulk Comptonization as outlined in \cite{begelman1987inverse}. All primed quantities are evaluated in the rest frame of the electron. The emergent Stokes parameters due to bulk Comptonization are
\begin{equation} \label{eqn:ics}
\begin{aligned}
    i' &= \frac{3}{16 \pi}\tau \int_{4 \pi}i_{0}'(\nu')(1+\cos^{2}w')d\Omega'(\rho',\alpha'), \\
    q' &= \frac{3}{16 \pi}\tau \int_{4 \pi}i_{0}'(\nu')(1-\cos^{2}w')\cos2\eta'd\Omega'(\rho',\alpha'), \\
    u' &= \frac{3}{16 \pi}\tau \int_{4 \pi}i_{0}'(\nu')(1-\cos^{2}w')\sin2\eta'd\Omega'(\rho',\alpha'),
\end{aligned}    
\end{equation} 
where $\nu'$ is the frequency of the incident photon and $i_{0}'$ is the intensity of the background radiation field, which we assume to be unscattered and unpolarized. It is transformed under the Lorentz boost as
\begin{equation} \label{eqn:frequency}
    i_{0}'(\nu') = i_{0}(\nu)\left(\frac{\nu'}{\nu}\right)^{3} = i_{0}(\nu)D_{1}^{3},
\end{equation}
and we take $i_{0}(\nu) = C(r)\nu^{-s}$ with $s = 1$. Here, $C(r)$ is a radius-dependent proportionality constant. Additionally, $\tau$ is the optical depth of the plasma, $\rho'$ and $\alpha'$ are the polar and azimuthal angles of the velocity vector of the incident photon, respectively, and $\Omega'(\rho',\alpha')$ is the solid angle spanned by these two angles. In the rest frame of the electrons, the $z$-axis coincides with the directional vector of the electron. Thus, the boost factor $D_{1} = \Gamma^{-1}(1 + \beta\cos\rho')^{-1}$. Furthermore, $w'$ is the angle between the velocity vector of the incident and the scattered photon, while $\eta'$ is the same angle but projected onto the tangent plane that is perpendicular to the direction of scattered photon,
\begin{equation}
    \cos\eta' = \frac{\cos\rho' - \cos\rho_{0}'\cos w'}{\sin\rho_{0}'\sin w'},
\end{equation}
so that $\rho_{0}'$ is the angle between the velocity vector of the electron and the scattered photon. In the Thomson regime, we have $\nu' = \nu_{0}'$, the scattered photon frequency, which transforms under Lorentz boosts as $\nu_{0}' = \nu_{0}\Gamma(1 - \beta\cos\rho_{0}) = \nu_{0}/D_{2}$, with $\beta = |v|/c$. Finally, the aberration formula yields
\begin{equation}
    \cos\rho_{0}' = \frac{\cos\rho_{0} - \beta}{1 - \beta\cos\rho_{0}},
\end{equation}
and it gives the angle in the rest frame. To transform back to the lab frame, we first multiply $i'$, $q'$, and $u'$ by the Lorentz boost factor
\begin{equation} \label{eqn:boost}
\begin{aligned}
    i &= i'(1 - \beta\cos\rho_{0})D_{2}^{3}, \\
    q &= q'(1 - \beta\cos\rho_{0})D_{2}^{3}, \\
    u &= u'(1 - \beta\cos\rho_{0})D_{2}^{3},
\end{aligned}    
\end{equation}
Equation \ref{eqn:ics} is defined so that the direction of motion of the electron coincides with the $y$-axis of the local polarization plane. We thus need to rotate the Stokes vector to account for the different orientations between the direction of motion of the electrons and the $y$-axis of the polarization plane of the observer
\begin{equation} \label{eqn:rotate}
\begin{aligned}
    q &\rightarrow q\cos2\xi - u\sin2\xi, \\
    u &\rightarrow q\sin2\xi + u\sin2\xi,
\end{aligned}    
\end{equation}
where $\xi$ is the required angle, which we will describe in detail in the next section.


\subsection{Radiative Transfer} \label{subsec:numerics-method}
\begin{figure}[htb!]
    \centering
    \includegraphics[width=1.0\linewidth]{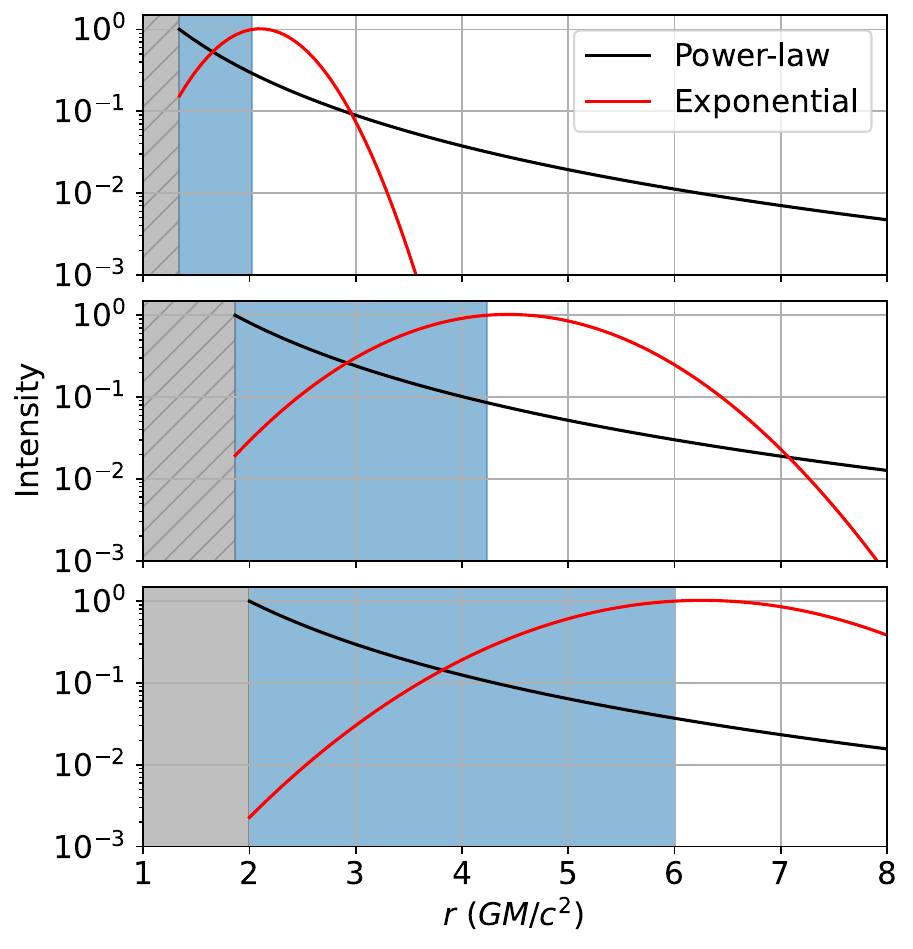}
    \caption{The functional form $C(r)$ of the background radiation field for the power-law model (Equation \ref{eqn:decrease}) and the exponential model (Equation \ref{eqn:increase}). We set the maximum of $C(r) = 1$. The grey region represents the space within the black hole horizon, while the blue region indicates the region within the ISCO.\label{fig:spectrum}}
\end{figure}
%

We now describe how to incorporate the bulk Comptonization formalism into a radiative transfer code. We use the polarized raytracing code \texttt{grtrans} \citep{dexter2009,dexter2010geokerr,dexter2016public,dexter2016grtrans} to compute polarimetric signatures. \texttt{grtrans} assumes the black hole is described by the Kerr metric, expressed in Boyer–Lindquist coordinates. To compute $\xi$, one would launch photons from the observer plane along geodesics to various points on the accreting plasma. One then solves for the Walker-Penrose constants \citep{walker1970quadratic} to compute the polarization basis $f^{\mu}$ at the plasma position. After this, one obtains $\xi$ by the vector product between the local polarization basis vector and $f^{\mu}$. One also rotates the locally computed Stokes $q$ and $u$ by $\xi$ and multiplies them by $g^3$, where $g$ is the redshift factor, to generate polarimetric signatures as seen by distant observers \citep{shcherbakov2011general}.  

The above procedures are typically carried out in the fluid frame, where the metric is locally Minkowski and the fluid $4-$velocity $u^{\hat{\mu}} = (1, 0, 0, 0)$. The local polarization basis is aligned such that the $y$-axis of the local polarization plane matches the spin axis of the accreting plasma \citep{agol1997effects}. In the framework of bulk Comptonization, however, one observes a background radiation field of a particular distribution (need not to be isotropic, see Section \ref{subsec:hyper}) in a specific frame where the plasma is moving with a relative $\Gamma$. We choose this frame to be the locally non-rotating frame \citep[ZAMO,][]{bardeen1972rotating}, which is also locally Minkowski, and the transformation matrix is given as
\begin{equation}
\begin{aligned}
    e^{\ (t)}_{\mu} &= (\sqrt{\frac{\Sigma\Delta}{A}}, 0, 0, 0), \\
    e^{\ (r)}_{\mu} &= (0, \sqrt{\frac{\Sigma}{\Delta}}, 0, 0), \\
    e^{\ (\theta)}_{\mu} &= (0, 0, \sqrt{\Sigma}, 0), \\
    e^{\ (\phi)}_{\mu} &= (-\frac{2ar\sin\theta}{\sqrt{\Sigma A}}, 0, 0, \sqrt{\frac{A}{\Sigma}}\sin\theta),
\end{aligned}
\end{equation}
where the constants are given as
\begin{equation}
\begin{aligned}
    \Sigma &= r^{2} + a^{2}\cos^{2}\theta, \\
    \Delta &= r^{2} - 2r + a^{2}, \\
    A &= (r^{2} + a^{2})^{2} - a^{2}\Delta \sin^{2}\theta,
\end{aligned}
\end{equation}
with $a$ is black hole spin, and $r$ the distance from the center of the black hole. We use brackets in the sub/superscript to indicate vector components in the ZAMO frame. Note that we set the speed of light $c$, black hole mass $M_{\rm BH}$, and gravitational constant $G$ to 1. The $4-$velocity, for example, transforms as $u^{(\nu)} = u^{\mu}e^{(\nu)}_{\mu}$. To summarize, we first transform the photon $4-$vector $k^{\mu}$, $u^{\mu}$, and $f^{\mu}$ as computed at the plasma position to the ZAMO frame. We then use Equations \ref{eqn:ics} -- \ref{eqn:boost} to compute $i$, $q$, and $u$. After this, we rotate the Stokes vector using Equation \ref{eqn:rotate}, with $\xi$ being the angle between $f^{(\mu)}$ and $u^{(\mu)}$. Finally, we multiply the Stokes vector by $g^{3}$, where $g = 1/k^{(t)}$, the time component of the photon $4-$vector in the ZAMO frame. Since the ZAMO frame is locally flat, one can compute $\Gamma$ and $\cos\rho_{0}$ as
\begin{equation}
\begin{aligned}
    \Gamma &= \frac{1}{\sqrt{1 - v_{(i)}v^{(i)}}}, \\
    \cos \rho_{0} &= \frac{k_{(i)}v^{(i)}}{\sqrt{v_{(j)}v^{(j)}}\sqrt{k_{(a)}k^{(a)}}},
\end{aligned}    
\end{equation}
where $i, j, a$ run from $r$ to $\phi$ and the $3$-velocity is given by $v^{(i)} = u^{(i)}/u^{(t)}$. Note that in using Equations \ref{eqn:ics} -- \ref{eqn:boost}, $\nu$ should be redshifted with respect to the observer at infinity. In GRRT, each light ray spans the same solid angle. Thus, the Stokes fluxes could be obtained as
\begin{equation}
\begin{aligned}
    I_{j,k} &= i_{y,k}\Delta\Omega, \\
    I_{\rm tot} &= \Sigma_{j}\Sigma_{k} I_{j,k},
\end{aligned}    
\end{equation}
where $j$ and $k$ are the indices along the horizontal and vertical axes of the camera, respectively, and $I_{j,k}$ is the flux for each pixel, while $I_{\rm tot}$ is the total flux. The Stokes $Q$ and $U$ fluxes are computed similarly. As we discussed, we cut off the bulk Comptonization physics beyond the ISCO, i.e., we set $i = i_{0}'$ and $q, u = 0$. To make the transition beyond the ISCO smooth, we employ the following transformation:
\begin{equation}
\begin{aligned}
    i &\rightarrow i[1 - \sigma\left( \frac{r - r_{I}}{r*} \right)] + i_{0}'\sigma\left( \frac{r - r_{I}}{r*} \right), \\
    q &\rightarrow q[1 - \sigma\left( \frac{r - r_{I}}{r*} \right)] , \\
    q &\rightarrow u[1 - \sigma\left( \frac{r - r_{I}}{r*} \right)] , 
\end{aligned}
\end{equation}
where $\sigma(x)$ is the logistic function ($f(x) = 1/[1 + \exp(-x)]$). Here, we choose $r^* = 0.1$.


\subsection{Model for the Background Radiation Field} \label{subsec:emission}
\begin{figure}[htb!]
    \centering
    \includegraphics[width=1.0\linewidth]{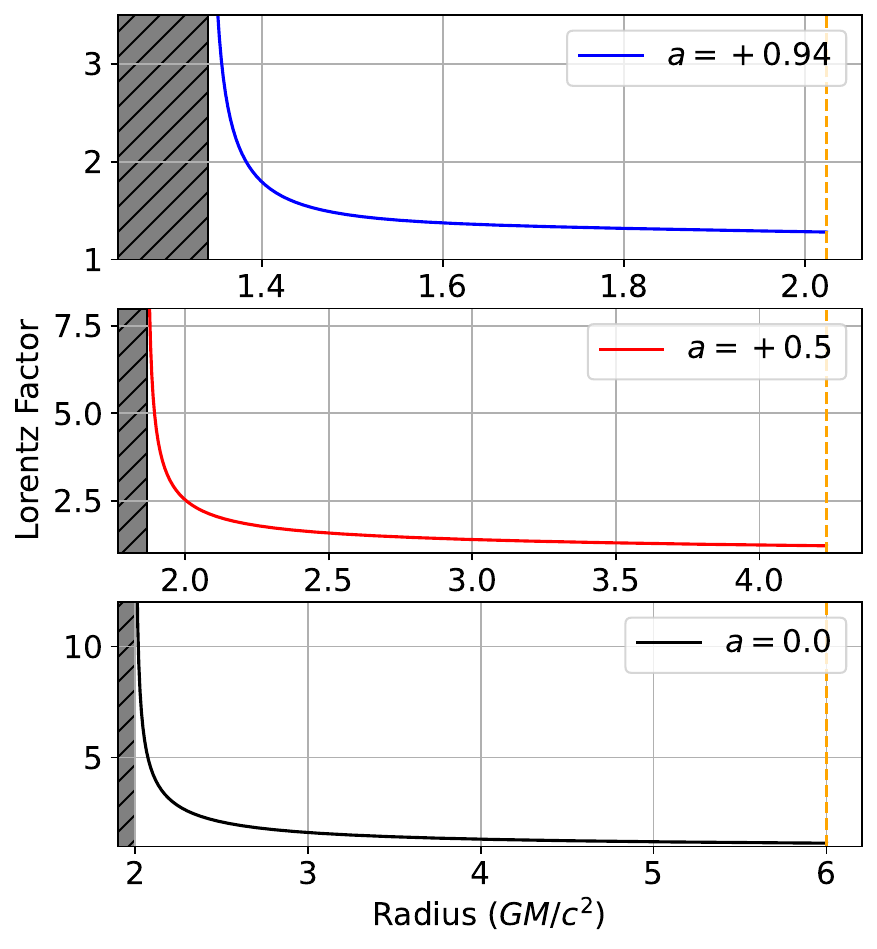}
    \caption{The Lorentz factor $\Gamma$ of the free-falling plasma as measured in the ZAMO frame for different spin $a$ of the black hole. The shaded region represents the space inside the event horizon. The vertical orange line represents the boundary of the plunging region. The Lorentz factor  diverges as $r \rightarrow r_{\rm BH}$. \label{fig:lorentz}}
\end{figure}
%

Next, we need to specify the functional form of $C(r)$. There is no commonly agreed-upon choice on the functional form of $C(r)$. We thus consider the simplest case, where $C(r)$ is a power-law of the radius
\begin{equation} \label{eqn:decrease}
    C \propto \left(\frac{r}{r_{\rm BH}}\right)^{-\alpha_{\gamma}},
\end{equation}
with $r_{\rm BH} = 1 + \sqrt{1 - a^{2}}$ the black hole outer horizon radius and $\alpha_{\gamma}$ the power-law index. Additionally, following \citet{cardenas2020modeling}, we employ a functional form of $C(r)$ where it increases from the horizon towards the ISCO and then declines afterward
\begin{equation} \label{eqn:increase}
    C \propto \left( \frac{r}{r_{I}} \right)\text{exp}\left(\frac{-(r - r_{I})^{2}}{r_{I} - \text{ln}[r_{I}] - 1}\right), 
\end{equation}
so that $r_{I}$ is the ISCO radius, which is given as 
\begin{equation}
\begin{aligned}
  r_{I} &= 3 + Z_{2} \pm \left((3 - Z_{1})(3 + Z_{1} + 2Z_{2})\right)^{1/2}, \\
  Z_{1} &= 1 + (1 - a^{2})^{1/3}\left( (1 + a)^{1/3} + (1 - a)^{1/3} \right), \\
  Z_{2} &= (3a^{2} + Z_{1}^{2})^{1/2},
\end{aligned}
\end{equation}
and the minus (plus) sign represents plasma on a prograde (retrograde) trajectory with respect to the spin of the black hole. We illustrate the functional forms of Equations \ref{eqn:decrease} and \ref{eqn:increase} in Figure \ref{fig:spectrum}.


\subsection{Model for Plasma within the Plunging Region} \label{subsec:plunging}
\begin{figure}[htb]
    \centering
    \includegraphics[width=1.0\linewidth]{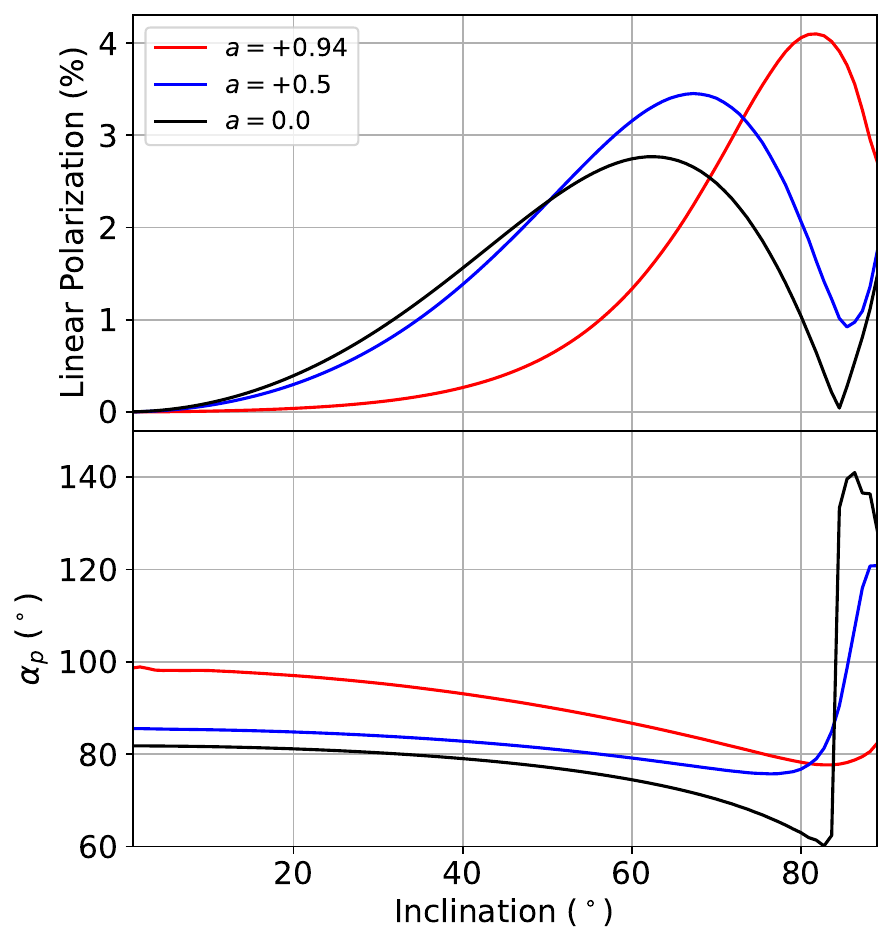}
    \caption{The spatially unresolved linear polarization (upper panel) and the spatially unresolved polarization angle (lower panel) against the observer inclination $\theta$ for models with different black hole spins. Here, we assumed the power-law background radiation field being isotropic in the ZAMO frame, and fixed $\tau = 1.0$ and $\alpha_{\gamma} = 3$. The linear polarization is sensitive to the variations in $a$, achieving a maximum of around $4\%$ when viewing at $\theta \sim 80^\circ$ for a rapidly spinning black hole. The polarization angle is not very sensitive to the variation of $\theta$ except for $\theta \sim 80 - 90$\,$^\circ$.\label{fig:lp_evpa_spin}}
\end{figure}
\begin{figure}[htb!]
    \centering
    \includegraphics[width=1.0\linewidth]{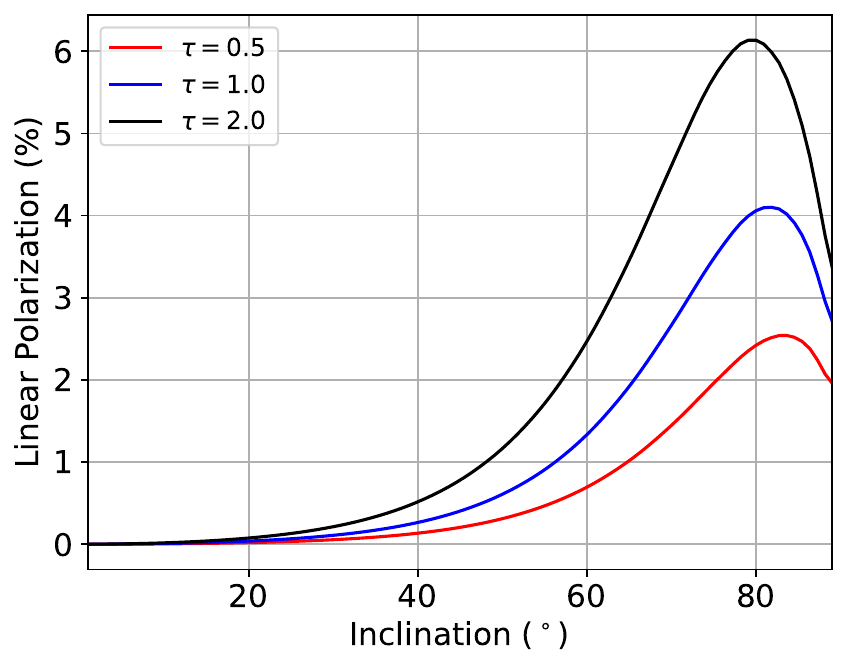}
    \caption{Same as Figure \ref{fig:lp_evpa_spin}, but for models with different optical depths within the plunging region. Note that we fixed $a  = +0.94$ and $\alpha_{\gamma} = 3$ instead. Increasing $\tau$ leads to an increase in the linear polarization. The polarization angle is independent of $\tau$ and thus is omitted.\label{fig:params_dep}}
\end{figure}
\begin{figure}[htb!]
    \centering
    \includegraphics[width=1.0\linewidth]{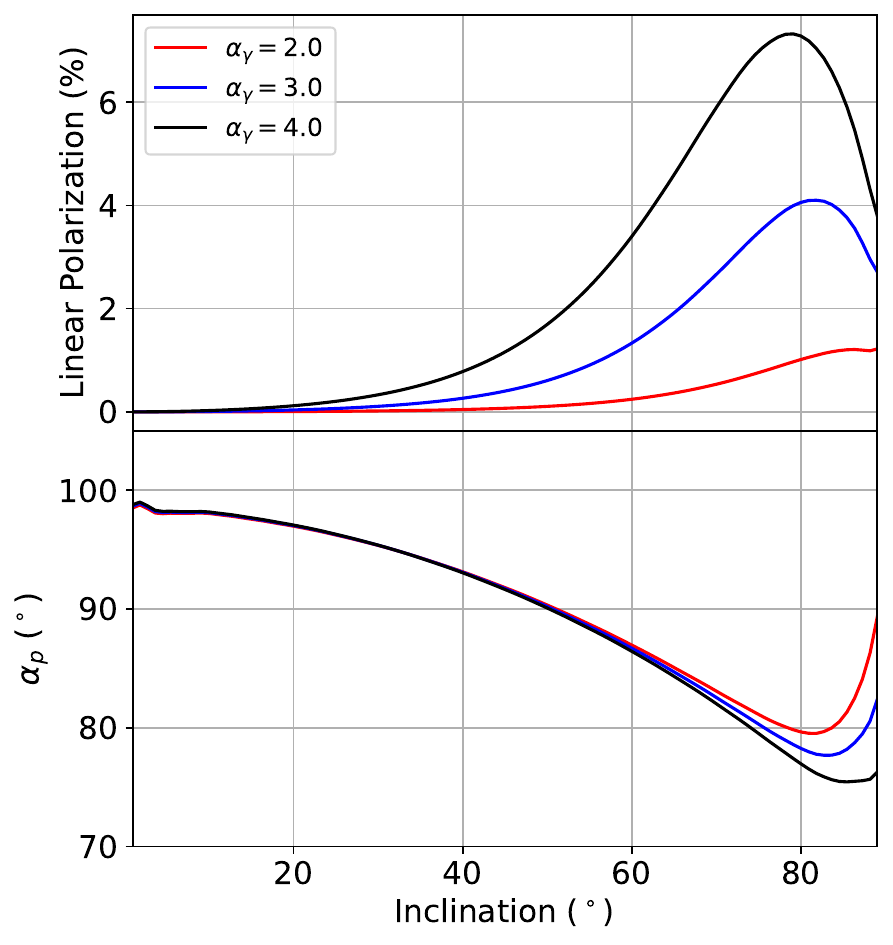}
    \caption{Same as Figure \ref{fig:lp_evpa_spin}, but for models with different $\alpha_\gamma$. Here, we fixed $\tau = 1.0$ and $a = +0.94$ instead. Increasing $\alpha_{\gamma}$ leads to an increase in the maximum linear polarization to around $7 - 8$\,$\%$ ($\alpha_{\gamma} = 4$). The polarization angle, however, is not very sensitive to the variation in $\alpha_{\gamma}$.\label{fig:params_alpha}}
\end{figure}
%

We now determine the bulk $4-$velocity profile of the plasma within the ISCO. We assume the plasma within the ISCO follows a geodesic \citep{cardenas2020modeling, dong2023x} for which the $4-$velocity is given by
\begin{equation} \label{eqn:geod}
\begin{aligned}
    u^{\mu} &= (u^{t}, u^{r}, 0, u^{\phi}), \\
    u^{t} &= \frac{1}{\Delta} \left[ \left( r^{2} + a^{2} + 2\frac{a^{2}}{r}\right)K - \frac{2aL}{r} \right], \\
    u^{r} &= -\sqrt{\frac{2}{3r_{I}}}\left(\frac{r_{I}}{r} - 1\right)^{3/2}, \\
    u^{\phi} &= \frac{1}{\Delta}\left[ \frac{2aK}{r} + \left(1 - \frac{2}{r} \right)L \right],
\end{aligned}
\end{equation}
where the energy $K$ and angular momentum $L$ are given by
\begin{equation}
\begin{aligned}
    K = \left(1 - \frac{2}{3r_{I}}\right)^{1/2}, \\
    L = 2\sqrt{3} \left(1 - \frac{2a}{3\sqrt{r_{I}}} \right).
\end{aligned}
\end{equation}
We illustrate the plasma Lorentz factor $\Gamma$ as measured in the ZAMO frame for the above mentioned models in Figure \ref{fig:lorentz}. Finally, the optical depth $\tau$ within the ISCO is assumed to be a constant. 


\subsection{Model Parameters} \label{subsec:modelparam}

Our toy accreting plasma model thus takes the following free parameters: \textbf{Hyperparameters} - the functional form of $C(r)$, which is either power-law (Equation \ref{eqn:decrease}) or exponential (Equation \ref{eqn:increase}), and the angular distribution of $i_{0}(\nu)$, which could be isotropic, perpendicular, or comoving (see Section \ref{subsec:hyper}); \textbf{Numerical parameters} - the plunging region optical depth $\tau$, the black hole spin $a$, and the power-law index $\alpha_{\gamma}$. Our raytracing spans a field of view of $50 \times 50$ (both in units of $GM_{\rm BH}/c^{2}$ measured with respect to the center of the black hole) and has an image resolution of $500 \times 500$ pixels while covering the observer inclination from $\theta = 0^\circ$ to $90^\circ$ with $100$ discrete angular grids.


\section{Results} \label{sec:results}


\subsection{Numerical Parameter Dependence} \label{subsec:paramdepend}
\begin{figure*}[htb!]
    \centering
    \includegraphics[width=1.0\linewidth]{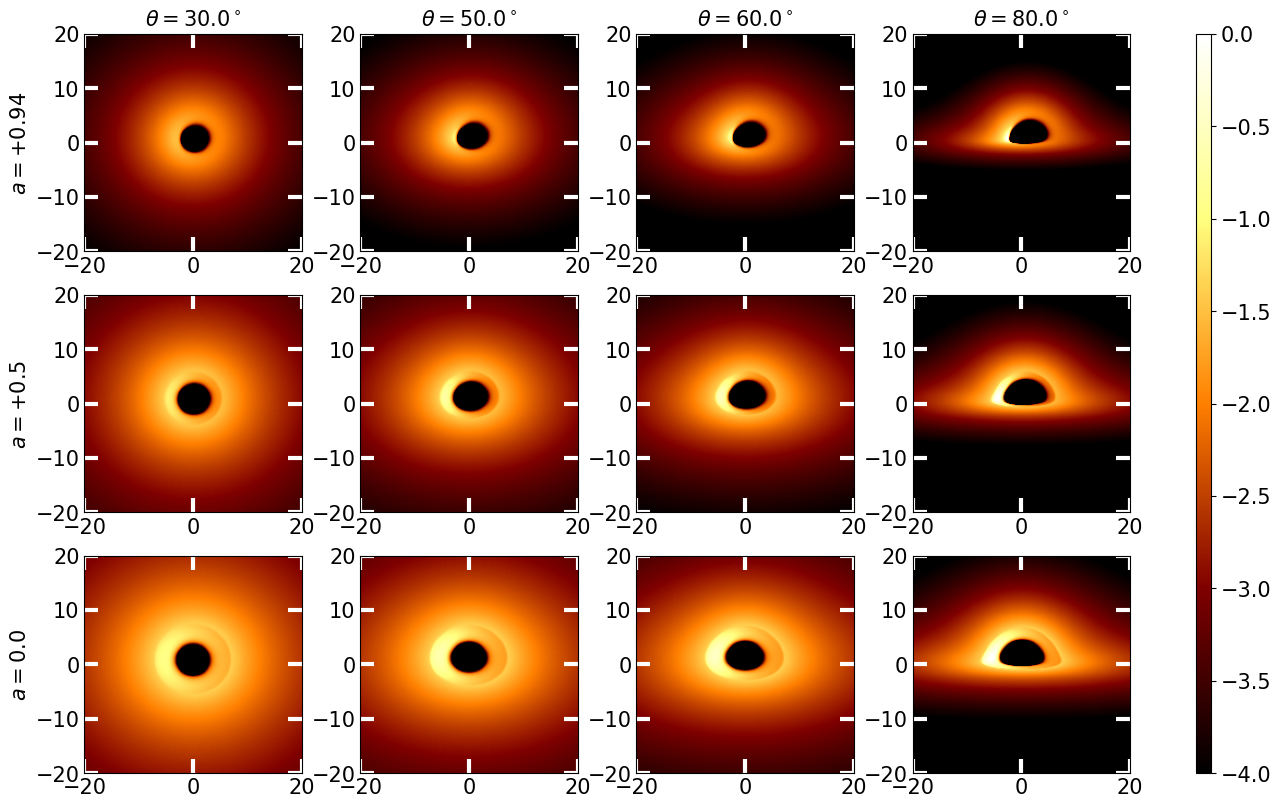}
    \caption{The Stokes $I$ images across different observer inclinations $\theta$ (shown at the top of each column) for models with varying spins (which are listed along the first column). Here, we assumed the power-law background radiation field being isotropic in the ZAMO frame, and fixed $\tau = 1.0$ and $\alpha_{\gamma} = 3$. Note that the fluxes are normalized to a range of $0$ to $1$ for each row separately. The normalized Stokes $I$ fluxes are in the log${10}$ scale, and we set a cut-off at $10^{-4}$. The combined effects of Doppler beaming and bulk Comptonization are manifested as emerging bright spots on the left-hand side of the black hole shadow. \label{fig:stokesi}}
\end{figure*}
\begin{figure*}[htb!]
	\centering
	\includegraphics[width=1.0\linewidth]{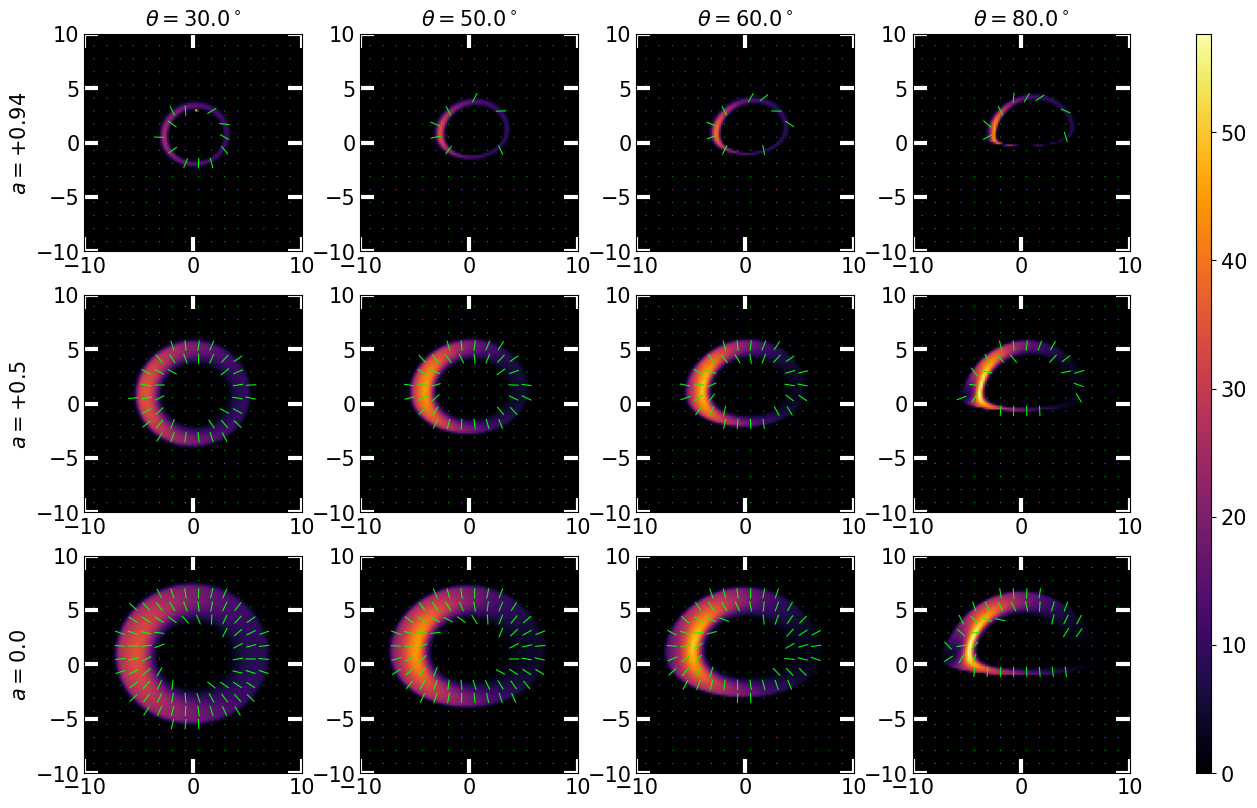}
	\caption{Same as Figure \ref{fig:stokesi}, but for the spatially resolved linear polarization map (in \%) with polarization ticks in green. All plots share a common color scale of percentage polarization. As with the normalized Stokes $I$ images, linear polarizations are more prominent on the left-hand side of the black hole shadow due to bulk Comptonization and Doppler beaming. The spatially resolved linear polarization can reach up to around $50$\,\%. Also, note that some polarization ticks turn vertical as $\theta$ increases, indicating photons launched with their $y$-component of the polarization basis being out of the plane. \label{fig:lp}}
\end{figure*}
\begin{figure*}[htb!]
	\centering
	\includegraphics[width=1.0\linewidth]{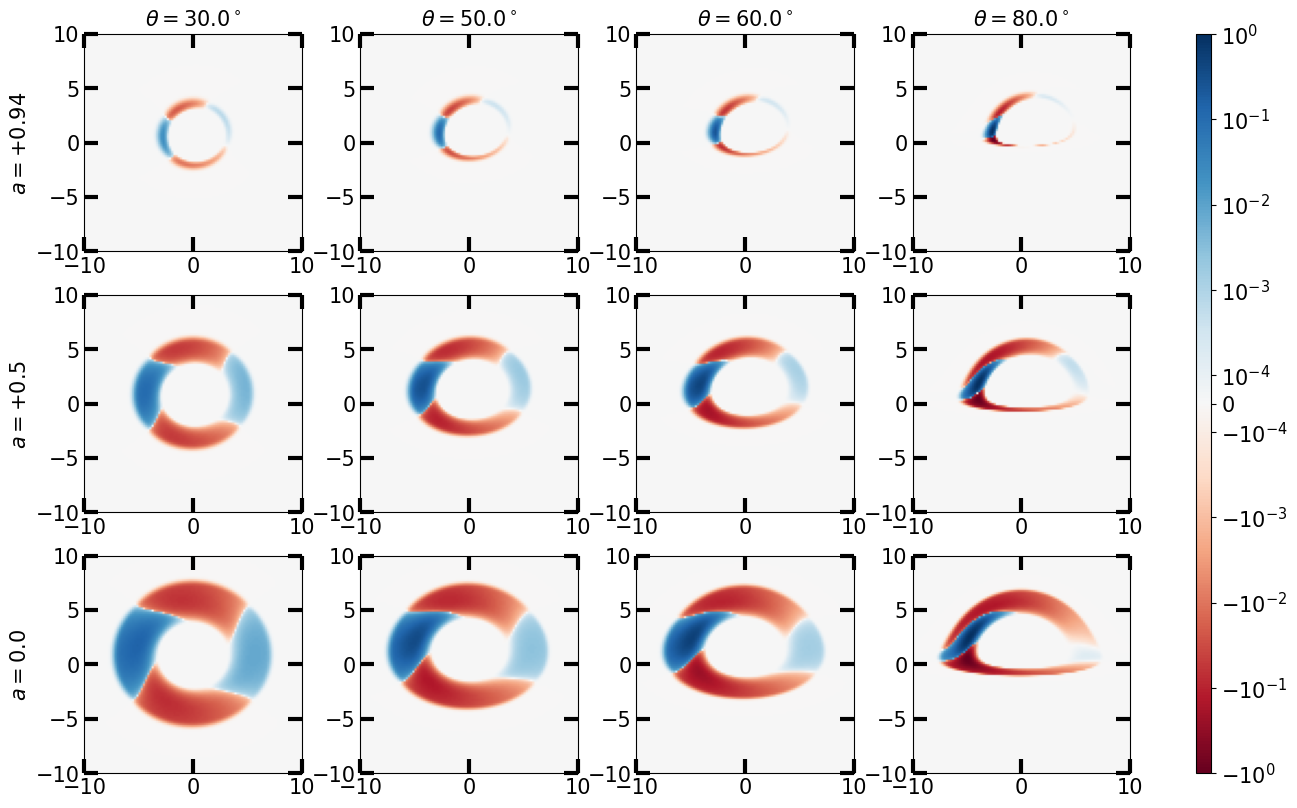}
	\caption{Same as Figure \ref{fig:stokesi}, but for the Stokes $Q$ fluxes. Note that the fluxes are normalized to a range of $-1$ to $1$ for each row separately. As $\theta$ increases, the positive Stokes $Q$ fluxes on the left-hand side of the black hole shadow are greater in magnitude due to Doppler beaming and bulk Comptonization but span a smaller solid angle. Negative Stokes $Q$ fluxes persist, have a smaller magnitude, but span a larger solid angle. The net effect is the substantial cancellation of the spatially unresolved Stokes $Q$ fluxes. \label{fig:stokesq}}
\end{figure*}

We first examine the parameter dependence of the spatially unresolved linear polarization on the black hole spin $a$, highlighting models that span $a = 0.0$, $+0.5$, and $+0.94$ while keeping $\tau = 1.0$ and $\alpha_{\gamma} = 3.0$. We assumed the background radiation field is a power-law and being isotropic in the ZAMO frame. We present the linear polarization against the observer inclination $\theta$ in the upper panel of Figure \ref{fig:lp_evpa_spin}. The curves are sensitive to variations in $a$. The peak linear polarization ($\sim$ 4\%) is achieved for $a = +0.94$ at $\theta \sim 80^\circ$, while the peak linear polarization is lower for black holes with smaller spin. Additionally, the angular location of the linear polarization peak shifts to smaller $\theta$ as $a$ decreases. Thus, for a given $\theta$, the linear polarization is non-monotonic with respect to $a$. We also present the polarization angle against the observer inclination $\theta$ in the lower panel of Figure \ref{fig:lp_evpa_spin} for the same set of models. Following \citet{begelman1987inverse}, we define the polarization angle $\alpha_{p}$ as
\begin{equation}
    \alpha_{p} = \frac{1}{2}\left[\pi - \text{tan}^{-1}\left(\frac{U}{Q}\right)\right], 
\end{equation}
and we find that the polarization angle stays very close to $90^\circ$ but varies around it. This indicates that the polarization angle is not very dependent on the observer inclination $\theta$, except perhaps at $\theta \sim 80 - 90^\circ$, which is due to the changing sign of the spatially unresolved Stokes $Q$/$U$ fluxes (see Section \ref{subsec:cancel}).

We then study similar set of models (ZAMO isotropic power-law), but we keep $a = +0.94$ and $\alpha_{\gamma} = 3.0$ while spanning $\tau = 0.1, 0.5,$ and $1.0$. We present the resulting linear polarization against the observer inclination $\theta$ in Figure \ref{fig:params_dep}. Since the scattered radiation and Stokes fluxes are proportional to $\tau$, an increase in $\tau$ increases the contribution of imprinted polarization from within the ISCO. Hence, the overall effect of increasing $\tau$ is scaling up the linear polarization. The polarization angle, however, is \textit{independent} (not shown) of the $\tau$ assumed because both $Q$ and $U$ are $\propto \tau$.

We also examine similar models (ZAMO isotropic power-law) that spans $\alpha_{\gamma} = 2, 3$, and $4$, while keeping $\tau = 1$ and $a = +0.94$ fixed. We present the linear polarization against the observer inclination $\theta$ in the upper panel of Figure \ref{fig:params_alpha}. We find that a steeper $C(r)$ changes the shape of the curve and increases the linear polarization, with a local maximum appearing and shifting to a smaller $\theta$. The value of the maximum increases to around $7 - 8$\,\% at $\alpha_{\gamma} = 4$. Such a trend presumably arises because, for a larger power index, scattering preferentially occurs closer to the black hole, where motion is more relativistic. Additionally, the higher power index reduces the contamination of unscattered radiation from outside the ISCO. The lower panel of Figure \ref{fig:params_alpha} shows $\alpha_{p}$ against $\theta$. We find that the polarization angle is not very sensitive to changes in $\alpha_{\gamma}$.

Our parameter study reveals that the polarimetry imprinted by the bulk Comptonization of relativistically moving plasma within the ISCO could be significant and depends on the black hole parameters and the plasma properties within the plunging region. We also note that the linear polarization against inclination angle has a generic shape across the varying numerical parameters — a higher linear polarization could be achieved when the black hole is viewed near edge-on.


\subsection{Image Morphology} \label{subsec:images}

We turn our attention to the spatially resolved Stokes fluxes to better understand the physics of bulk Comptonization and the numerical parameter dependencies. We first focus on the Stokes $I$ fluxes (showing $I_{j,k}$ as an image) and present the image morphology with varying black hole spins in Figure \ref{fig:stokesi}. We concentrate on models with the power-law background radiation field that is isotropic in the ZAMO frame, keeping $\tau = 1.0$, $\alpha_{\gamma} = 3.0$, while spanning $\theta = 30, 50, 60,$ and $80^\circ$. When $\theta$ is small, the Stokes $I$ images are mostly azimuthally symmetric. As $\theta$ increases, the Stokes $I$ fluxes on the left-hand side of the black hole shadow stand out due to 1) Doppler beaming and 2) bulk Comptonization. This is because most photons are now launched from within the ISCO. Hence, photons received by the camera are more likely to be up-scattered by the relativistically moving plasma. The solid angle spanned by the prominent Stokes $I$ fluxes increases as $a$ decreases because the ISCO gets larger when the black hole spin is smaller.

Next, we present the spatially resolved polarization maps in Figure \ref{fig:lp}, overlaying the maps with ticks to indicate the spatially resolved polarization angle. When $\theta$ is small, the polarization ticks are almost radial with respect to the center of the image but become twisted if the ticks are too close to the black hole. This occurs because the spatially resolved polarization angle is perpendicular to the velocity field of the plasma. Away from the horizon (but still inside the ISCO), the plasma velocity is almost azimuthal, but the radial component starts dominating when the plasma gets too close to the horizon, twisting the polarization ticks. As $\theta$ increases, some polarization ticks turn vertical. These signals are contributed mainly by photons that are away from the black hole horizon. The $y$-component of the polarization basis of these photons is out of the plane, making an angle of $90^\circ$ with respect to the velocity of the plasma.

The resolved linear polarization is greatest near the black hole shadow because the linear polarization imprinted by bulk Comptonization strongly depends on the Lorentz factor $\Gamma$ of the plasma, which increases substantially as $r$ approaches the event horizon. We note that the linear polarization morphology is identical across varying $\alpha_{\gamma}$. It may seem counterintuitive that the functional form of $C(r)$ does not affect the spatially resolved linear polarization. However, note that $q' \propto C$ and $i' \propto C$, so the linear polarization $q'/i'$ is therefore \textit{independent} of $C$ and solely depends on $\Gamma$.


\subsection{Stokes Flux Cancellations} \label{subsec:cancel}


Figure \ref{fig:lp} shows that the spatially resolved linear polarization is large and could approach around $50$\,\%, but we find from Section \ref{subsec:paramdepend} that the maximum linear polarization is around $7 - 8$\,\% for a very steep function of $C(r)$. One apparent reason for the drastic difference is the dilution from the disk background radiation field. Another, less obvious reason is the cancellation of Stokes fluxes.

To understand this, we show maps of the normalized Stokes $Q$ fluxes in Figure \ref{fig:stokesq}. The model parameters are the same as those in Figure \ref{fig:stokesi}. When $\theta$ is small, the Stokes $Q$ fluxes are periodic and alternating in sign, which is due to the intrinsic azimuthal symmetry of the model. Accordingly, the total Stokes $Q$ fluxes sum to approximately zero. However, as $\theta$ increases, the positive Stokes $Q$ fluxes on the left-hand side of the black hole shadow increase in magnitude due to 1)stronger Doppler beaming and 2) bulk Comptonization. Despite this, the positive Stokes $Q$ fluxes cover a smaller solid angle. Meanwhile, we find that negative Stokes $Q$ fluxes persist. Even though they are weaker in magnitude, they span a larger solid angle than the positive ones. These negative Stokes $Q$ fluxes largely cancel out the positive Stokes $Q$ fluxes, causing the net linear polarization to be smaller than the resolved one.

We also find such alternate sign cancellation for the Stokes $U$ fluxes (not shown), and this effect persists for different spins of the black hole. However, we find that the solid angle spanned by the positive/negative Stokes $Q$/$U$ fluxes can vary across different values of numerical parameters, and hence, the net Stokes $Q$/$U$ fluxes differ accordingly. This explains why the net linear polarization depends on the varying numerical parameters. We note that the origin of the negative Stokes $Q$/$U$ fluxes can be traced via the polarization ticks in Figure \ref{fig:lp}. Comparing it with Figure \ref{fig:stokesq}, we find that negative Stokes $Q$ fluxes are always associated with a local polarization angle larger than $45^\circ$ (measured with respect to the horizontal axis). Thus, these are photons launched with a $y$ polarization basis that makes an angle larger than $45^\circ$ with the local velocity vector.


\subsection{Hyperparameter Dependence} \label{subsec:hyper}
\begin{figure}[htb!]
	\centering
	\includegraphics[width=1.0\linewidth]{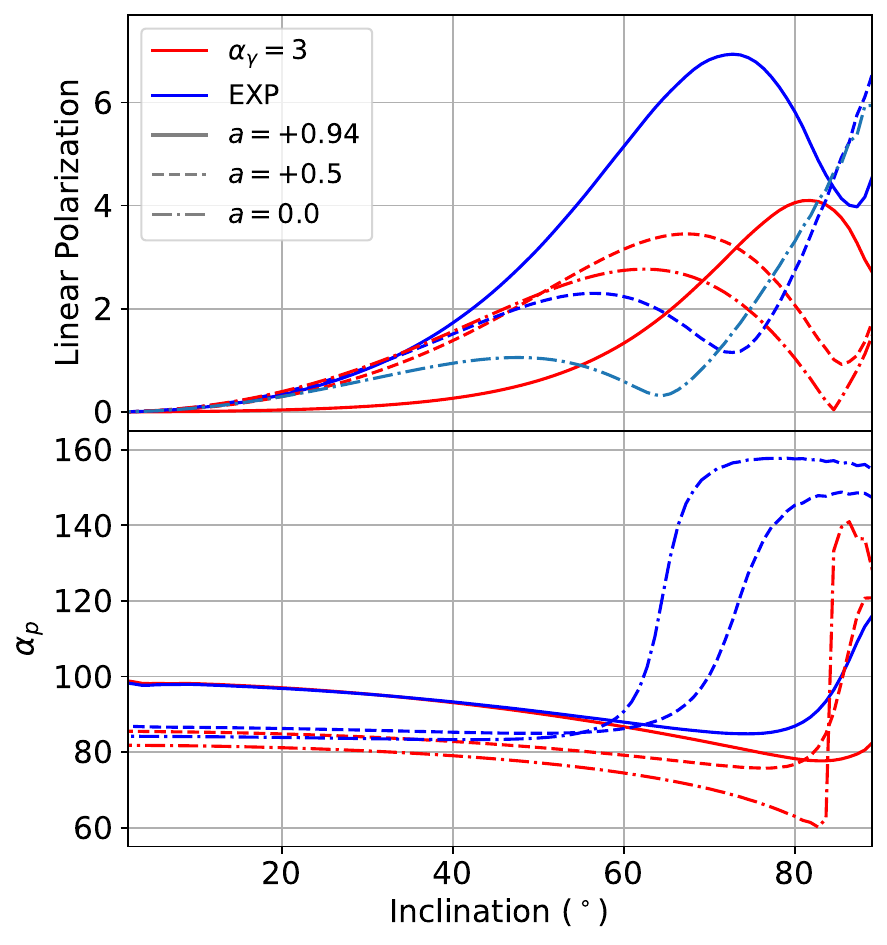}
	\caption{Comparison between models with the power-law (red lines) and the exponential (blue lines) background radiation field (both being ZAMO isotropic) across different black hole spins (shown as different linestyles). Here, $\alpha_{\gamma} = 3$ for the power-law model, and $\tau = 1.0$ for both. We show the linear polarization (upper panel) and polarization angle (lower panel) against observer inclination $\theta$. The results are quite similar, except that the linear polarization for the exponential model with $a = +0.94$ is scaled up, achieving a maximum value of approximately $7 - 8$\,\%, and the peak is shifted to a smaller $\theta$. \label{fig:lpevpa_decay}}
\end{figure}
\begin{figure}[htb!]
	\centering
	\includegraphics[width=1.0\linewidth]{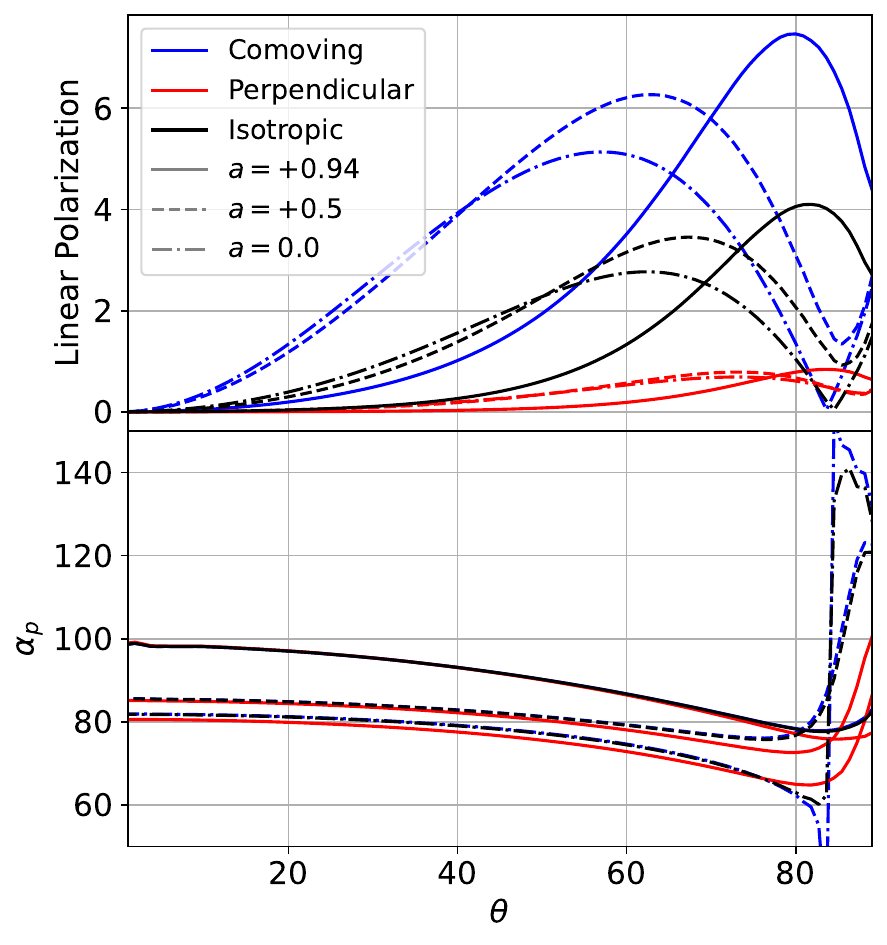}
	\caption{Similar to Figure \ref{fig:lpevpa_decay}, but here we compare models with and without anisotropy in the background radiation field (both assumed to be a power-law) across different black hole spins. Here, the comoving, perpendicular, and isotropic models are listed as blue, red, and black lines, respectively. We fixed $\tau = 1.0$ and $\alpha_{\gamma} = 3$. The comoving (perpendicular) model gives more (less) polarized due to enhanced (reduced) head-on collisions with the relativistic electrons. \label{fig:aniso}}
\end{figure}
\begin{figure*}[htb!]
    \centering
    \gridline{
    \fig{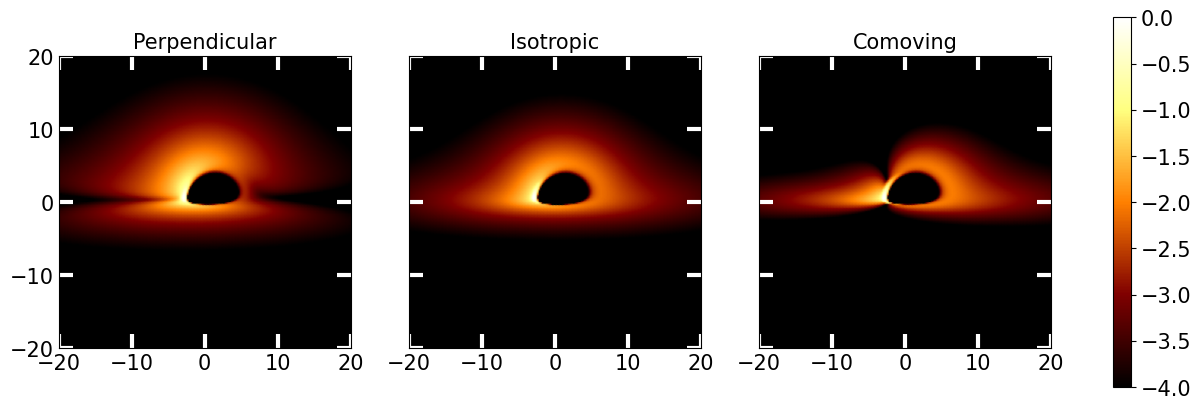}{1.0\textwidth}{(a)}}
    \gridline{
    \fig{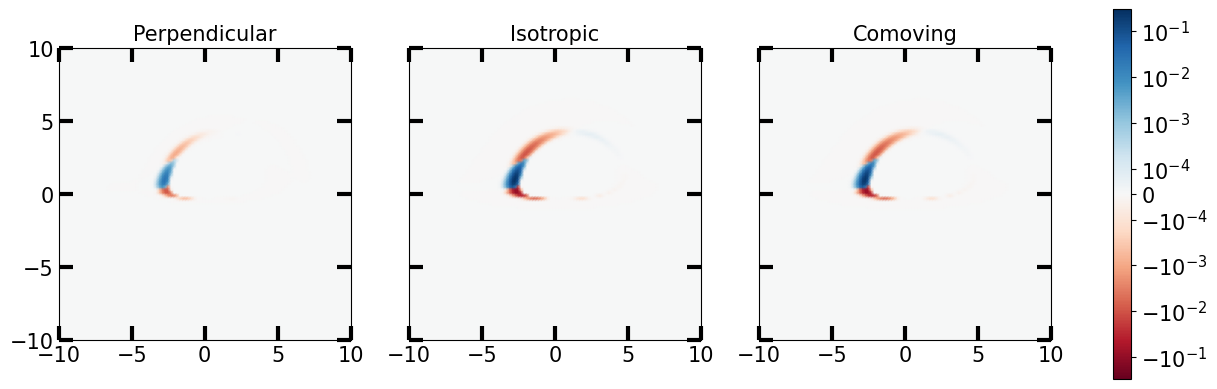}{1.0\textwidth}{(b)}}
    \caption{Comparison of models with and without anisotropy (listed at the top of each image) in the background radiation field, in terms of Stokes $I$ (a) and Stokes $Q$ (b) images. The black hole is viewed at $80^\circ$, fixed $\tau = 1.0$, and assumed the power-law background radiation field with $\alpha_{\gamma} = 3$. In (a), we normalize the Stokes $I$ images by their domain-wise maximum independently, while in (b), we normalize the Stokes $Q$ images by the maximum computed in (a), also independently. \label{fig:aniso_image}}
\end{figure*}

Here, we examine if the results we obtained in Section \ref{subsec:paramdepend} are sensitive to the hyperparameters of the model, namely the distribution $C(r)$ and the angular distribution of $i_{0}(\nu)$. We first focus on the functional form of $C(r)$, comparing models with the power-law and the exponential background radiation field, where the former (latter) has a functional form of $C(r)$ that decreases (increases) from the event horizon towards the ISCO (see Figure \ref{fig:spectrum}). We also fixed $\alpha_{\gamma} = 3.0$ for the power-law model and $\tau = 1.0$ for both models, and assumed that the radiation is isotropic in the ZAMO frame. The results of the polarimetry across black hole spins are shown in Figure \ref{fig:lpevpa_decay}. The variations of the linear polarization against the observer inclination $\theta$ are quite similar across the two models, except that for model with $a = +0.94$, the overall magnitudes are scaled up and the peak of the linear polarization (approximately $7 - 8$\,\%) moves to a smaller $\theta$ ($\sim 70^\circ$). The polarization angle is also similar between the two models. We note that the amount of polarization imprinted depends on the magnitude of the bulk velocity, which diverges as $r \rightarrow r_{\rm BH}$. However, electromagnetic emission coming from $r$ close to $r_{\rm BH}$ also experiences a significant amount of redshift. Therefore, the functional form of Equation \ref{eqn:increase} selects an emission radius where the bulk velocity is large but the photons are not heavily redshifted, explaining the increase in the polarization magnitudes. Nonetheless, we conclude that the magnitudes of the linear polarization depend on the functional form of $C(r)$, an important parameter for constraining the background radiation intensity profile of the accreting plasma, especially within the ISCO.

Unlike \citet{begelman1987inverse} and \citet{dexter2024relativistic}, however, we relax the assumption that the background radiation field is isotropic (in the ZAMO frame). We recall that the physics of bulk Comptonization is the introduction of a radiation anisotropy in the rest frame of the electron via the relativistic frequency shift (cf. Equation \ref{eqn:boost}). Thus, the electron sees an anisotropic radiation field regardless of the isotropy of the radiation field in the ZAMO frame. We call this the \textit{relativistic} anisotropy. If, however, the background radiation field to be reprocessed by bulk Comptonization is initially anisotropic, it would then add an extra angular dependence to the background radiation intensity profile (which could also be radially dependent). We call this the \textit{ab initio} anisotropy. For simplicity, we model the ab initio anisotropy by assuming the proportionality function $C(r)$ is replaced by a product of two functions, $A(r)B(\rho')$, with $A(r) \propto r^{-\alpha_{\gamma}}$, thereby introducing the anisotropy as an extra multiplicative factor to the background radiation intensity profile. Note that such an assumption means that the anisotropy has no radial dependence, and the background radiation field is symmetric along $\alpha'$, which might not hold in general but should be fairly good for our first-order estimation. Also, we specify the angular dependence in the rest frame of the electron rather than in the ZAMO frame without any loss of generality because it is equivalent to first specifying the anisotropy in the ZAMO frame and then transforming all the angular coordinates from the ZAMO to the rest frame of the electron. We consider two functional forms of $B(\rho')$: $B(\rho') \propto \text{sin}^{2}\rho'$ (the perpendicular model) and $B(\rho') \propto \text{cos}^{2}\rho'$ (the comoving model), where the former (latter) represents the background radiation field as concentrated along the direction of (perpendicular to) the velocity of the electron.

We first show the resulting spatially unresolved linear polarization and polarization angle against the observer inclination in the upper panel of Figure \ref{fig:aniso}. Note that the spin of the black hole is also varied. We find that including radiation anisotropy impacts the linear polarization. If the background radiation field is more concentrated along the electron's direction of motion, the maximum linear polarization increases, and its peak shifts towards a smaller $\theta$. The opposite is true for a background radiation field that is more perpendicular to the electron's velocity vector. This is because photons that are concentrated along the electron's direction of motion are more likely to collide head-on with the relativistic electron, thus imprinting polarization. We note that the polarization angle does not change appreciably with different $B(\rho')$.

We also show the Stokes $I$ and $Q$ image morphologies of models with different radiation anisotropies in Figure \ref{fig:aniso_image}. We take $\theta = 80^\circ$, fixed $\tau = 1.0$ and $a = +0.94$, and assumed the background radiation field is a power-law with $\alpha_{\gamma} = 3.0$. We find that assuming different angular dependencies of $B(\rho')$ changes the image morphology: the Stokes $Q$ images show smaller magnitudes for the perpendicular model due to the lack of head-on collisions, while the Stokes $I$ images of the perpendicular model span a larger solid angle than those of the isotropic and comoving models. This is because photons are launched with a directional vector that makes some angle with the mid-plane to reach the observer, and the electron's velocity lies along the mid-plane. The perpendicular (comoving) model thus helps enhance (reduce) the Stokes $I$ signals. Also, note that the normalized Stokes $Q$ magnitudes are very similar between the isotropic and comoving models. Thus, the enhanced linear polarization for the comoving model is primarily due to the reduction of disk dilution. We remark that the imprinted polarization signal could be further enhanced if the background radiation field is even more highly collimated along the electron's direction of motion.  The Stokes $U$ images are very similar to those of Stokes $Q$ and are not shown.


\section{Discussion} \label{sec:discuss}


\subsection{Comparsion to Thermal Scattering} \label{subsec:thermal}
\begin{figure}[htb!]
	\centering
	\includegraphics[width=1.0\linewidth]{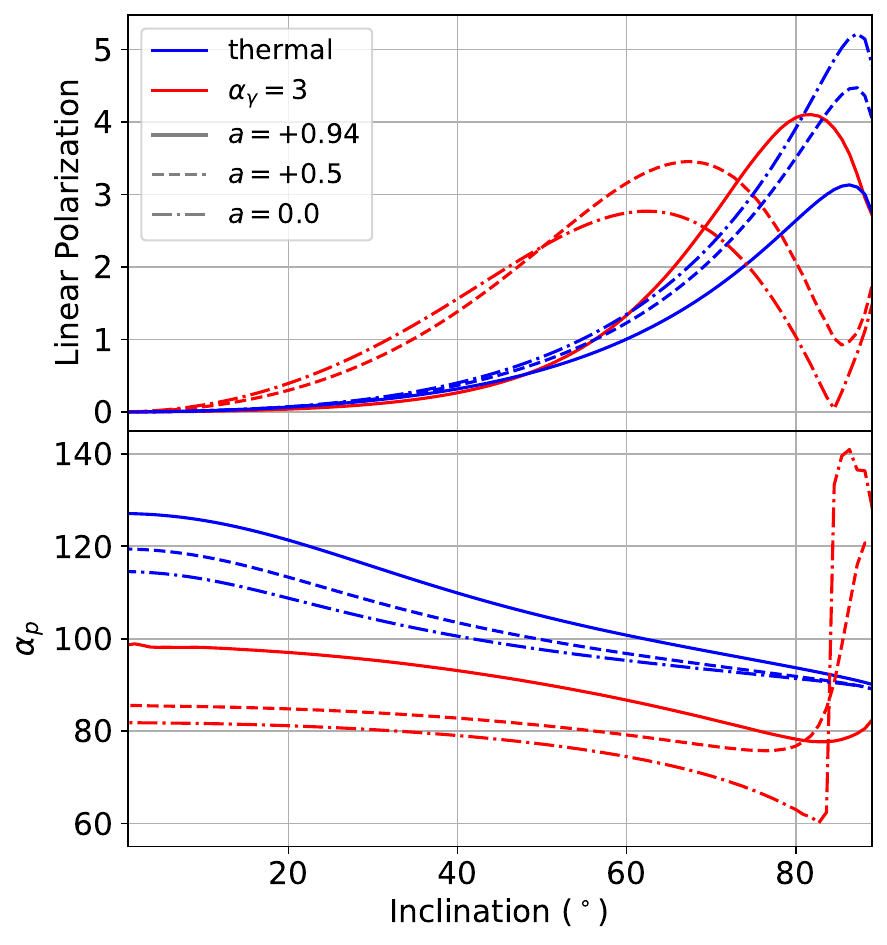}
	\caption{Similar to Figure \ref{fig:lpevpa_decay}, but comparing our toy accreting plasma model (red lines) with the Novikov-Thorne thin disk (blue lines) across different black hole spins. For our toy model, we assumed the (ZAMO isotropic) power-law background radiation field, and fixed $\tau = 1.0$ and $\alpha_{\gamma} = 3$. For the NT thin disk, the black hole mass is $10\,M_{\odot}$ and the mass accretion rate is $0.1$ of the Eddington limit. For a rapidly spinning black hole, the polarization imprinted by bulk Comptonization within the ISCO can have a maximum value larger than that produced by thermal scattering. \label{fig:thermal}}
\end{figure}

Photons experience thermal electron scattering that imprints polarimetric signatures, which could contaminate those from the bulk comptonization inside the ISCO. To obtain a qualitative understanding of how the two compete with each other, we compare the polarimetric signatures imprinted by bulk Comptonization with those by thermal scattering. To compute the polarization imprinted by thermal scattering, we adopt the Novikov-Thorne thin disk \citep{novikov1973astrophysics} model, assuming the black hole mass is $10\,M_{\odot}$ and the accretion rate is at $0.1$ of the Eddington limit ($\dot{M}_{\rm Edd} = 4\pi G M_{\rm BH}m_{p}/\sigma_{t}/c$), also assuming the Chandrasekhar thermal scattering model \citep{chandrasekhar2013radiative}. We compare the linear polarization with our toy model, for which we fixed $\tau = 1.0$ and assumed the background radiation field (isotropic in the ZAMO frame) is a power-law with $\gamma = 3.0$. Results across black hole spins are shown in Figure \ref{fig:thermal}. For a small spin $a$, the linear polarization imprinted by bulk Comptonization is larger than that due to thermal scattering when $\theta$ is small, but the maximum value of the former is smaller than the latter. For a rapidly spinning black hole, however, the maximum value of the linear polarization produced by bulk Comptonization can exceed that of thermal scattering. We also find that $\alpha_{p}$ in the thermal scattering model does not exhibit sharp changes at large $\theta$ as bulk Comptonization does. This is because the thermal scattering model does not suffer from the substantial Stokes flux cancellations. Finally, the resultant linear polarization when the two effects are combined depends on the plasma model. However, given that the imprinted linear polarization of the two processes is similar in magnitude, we expect that the inclusion of thermal scattering will produce order-of-unity changes (in terms of \%) to our current results.


\subsection{Flux Contribution from within the ISCO} \label{subsec:iscofluxes}
\begin{figure}[htb!]
    \centering
    \includegraphics[width=1.0\linewidth]{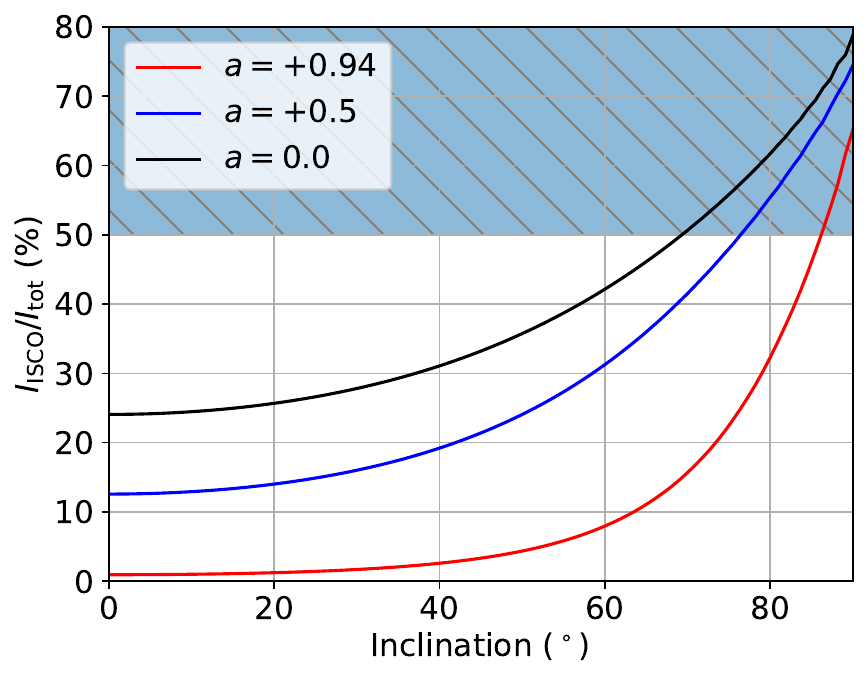}
    \caption{Ratio of the observed radiation flux from within the ISCO to the total from all plasma (in \%) across different black hole spins. Here, we assumed the (ZAMO isotropic) power-law background radiation field, and fixed $\tau = 1.0$ and $\alpha_{\gamma} = 3.0$. The shaded region represents flux from within the ISCO being more than $50$\,\% of the total flux.  Model with $a = +0.94$ has an flux from within the ISCO contributing only about $25 - 30$\,\% of the total flux when the black hole is viewed at around $80^\circ$, but the linear polarization is approximately $4\,$\%. Thus, it is possible that the polarimetric signatures from bulk Comptonization are significant even if the observed flux is not heavily peaked towards the plunging region. \label{fig:reflection}}
\end{figure}

The linear polarization imprinted by bulk Comptonization within the ISCO is diluted by the disk radiation. Thus, for the bulk Comptonization to be significant, the background radiation field needs to increase towards the plunging region. After being reprocessed by bulk Comptonization, the total radiation flux received by the observer will be peaked towards the plunging region. The assumption that the observed radiation flux is peaking heavily towards the plunging region might be too strict and may not be consistent with observations of BHBs in their soft state or of luminous AGN. To assess the likely astrophysical relevance of our bulk Comptonization models, we compute the ratio of the Stokes $I$ flux being reprocessed by bulk Comptonization from within the ISCO, $I_{\rm sc}$, to that of the total (within plus beyond the plunging region), $I_{\rm tot}$, i.e., $I_{\rm sc}/I_{\rm tot}$, where $I_{\rm tot} = I_{\rm sc} + I_{\rm eisco}$. We focus on models with the (ZAMO isotropic) power-law background radiation field, and fixed $\tau = 1.0$ and $\alpha_{\gamma} = 3.0$. Results across different black hole spins are shown in Figure \ref{fig:reflection}. The flux from within the ISCO could exceed half of the total flux when the black hole is viewed edge-on. This ratio is larger for black holes with smaller spins, probably because of the larger size of the ISCO. Surprisingly, we find that model \textit{g\_i\_p\_p3\_a094\_t10} viewed at around $80^\circ$ has a significant linear polarization (around $4$\,\%) but not a considerable percentage of flux (around $25 - 30$\,\%) from within the ISCO. Thus, we suggest that some of our models with considerable linear polarization could be astrophysically relevant.


\section{Conclusions} \label{sec:conclu}


The detection of X-ray polarization in a number of BHBs by  {\it IXPE} has rekindled interest in theoretical X-ray polarization physics and could help us better understand plasma under extreme astrophysical conditions. Additionally, there is increasing interest within the field in examining electromagnetic emissions from within the plunging region, which could potentially be used to probe physics under the strong gravity regime. However, the polarization signatures of gas within the ISCO is largely unexplored. Given this, we present a toy accreting plasma model to determine the polarimetric emission due to bulk Comptonization within the ISCO. The toy model consists of relativistically moving plasma that has a constant optical depth (within the ISCO) and emits power-law spectrum radiation. The radiation is reprocessed by bulk Comptonization within the ISCO, imprinting polarization onto it. The effects of strong gravity are captured via a GRRT code. Our main findings are summarized as follows.
\begin{enumerate}
    \item The linear polarization imprinted by bulk Comptonization is sensitive to the black hole and plasma parameters. By fine-tuning some of the free numerical parameters, we find that the net linear polarization could be as large as around $7 - 8$\,\% when the black hole is viewed near edge-on. Even if the observed radiation flux is not heavily peaked towards the plunging region, the linear polarization could be significant (around $4$\,\%).
    
    \item The spatially resolved (around $50$\,\%) and unresolved linear polarization differ significantly from each other. This is due to 1) the substantial cancellations of the Stokes $Q$ and $U$ fluxes of alternating signs and 2) dilution from the disk radiation. Nonetheless, the resultant spatially unresolved linear polarization is approximately $O(1)$\,\%, reflecting the importance of bulk Comptonization in imprinting polarization.
    
    \item The alternating signs of the Stokes $Q$ and $U$ fluxes are due to photons launched with their $y$-polarization basis making different angles with the plasma velocity, which are parallel-transported and rotated to give either positive or negative signs. This is an intrinsically general relativistic effect. The resultant magnitudes and signs depend on the black hole and plasma parameters.

    \item We relax the assumption that the radiation field be isotropic before being reprocessed by bulk Comptonization, and find that introducing radiation anisotropy impacts the linear polarization. If the radiation field is more aligned (perpendicular) to the electron's velocity, the resulting linear polarization is larger (smaller).
    
    \item The linear polarization imprinted by bulk Comptonization from within the ISCO is comparable to or even larger than that due to thermal electron scattering in a standard Novikov-Thorne thin disk. If both effects are to be included, we expect an order unity change change to our current results. We leave this for a future study.
    
    \item The linear polarization imprinted by bulk Comptonization is maximized when the unresolved signal is dominated by a spatially small region (effectively making it resolved), i.e., a steep radial background intensity profile around a rapidly spinning black hole, plus a marginally optically thin plasma within the ISCO.
\end{enumerate}

We remark that our toy accreting plasma model is relatively simple, and some of the assumptions we made might not hold in general. For instance, we expect that the background radiation field would be polarized before being bulk Comptonized if the source of emission is, for instance, synchrotron radiation or scattered by a disk corona; this might change the final results. We also assume a \textbf{temporally} and \textbf{spatially} constant optical depth within the ISCO. To address the temporal variation, we include models with different constant $\tau$ so one can interpolate between different $\tau$ to mimic the temporal variation; in reality, we expect that the optical depth will decrease from the ISCO to the event horizon, but not too rapidly\footnote{See, for example, Figure 1 in \citet{hankla2022non}.}. So the spatially constant $\tau$ assumption should be fairly good for our toy accreting plasma model. The radial dependence of the background radiation field, $C(r)$, is still quite arbitrary and could be calculated \textbf{ab initio} via frequency-dependent radiative transfer, which is beyond the scope of our current study. Finally, the plasma might not be exactly freely falling. In magnetically dominated accretion, for example, the accumulation of the horizon-penetrating flux could reduce the inflowing velocity of the plasma. Some of these weaknesses could be overcome by more sophisticated modeling, e.g., radiative GRMHD simulations of BHBs. 

We make two additional points in closing. First, past studies on linear polarization from BHBs have found significant changes in polarization due to returning radiation \citep{agol2000magnetic, schnittman2009x}. Our main focus is to explore the new effect of bulk Comptonization and quantify it relative to other sources of linear polarization, but returning radiation would likely change our results in a similar way as for thermal disk emission. Second, since we assumed the background radiation field obeys a power-law spectrum, the imprinted linear polarization is independent of the frequency. This can be easily seen when one substitutes Equation \ref{eqn:frequency} into Equation \ref{eqn:ics}, giving $q'$ and $u' \propto \nu^{-s}$ even after the Stokes vectors are rotated. These two points might be the main caveats of our toy model, but they could be improved by more realistic modeling. Ultimately, by combining all commonly discussed plasma processes with bulk Comptonization, one can quantify the importance of bulk Comptonization and test whether it could be one piece of the missing physics for the unexpectedly high linear polarizations of Cygnus X-1 and 4U 1630-47. We leave this work for the future.


\begin{acknowledgments}
We thank the referee for providing thoughtful and constructive comments that have helped us improve the paper.  Ho-Sang (Leon) Chan acknowledges support from the Croucher Scholarship for Doctoral Studies by the Croucher Foundation. MB and JD acknowledge support from NASA Astrophysics Theory Program grants 80NSSC22K0826, 80NSSC24K1094, and IXPE Guest Observer program grant 80NSSC24K1174.
\end{acknowledgments}


\bibliography{main}{}
\bibliographystyle{aasjournal}

\end{document}